# Empirical physical formula for potential energy curves of $^{38-66}$Ti isotopes by using neural networks

S. Akkoyun[a], T. Bayram[b], S.O. Kara[c], N. Yildiz[a]

[a]Department of Physics, Cumhuriyet University, Sivas, Turkey

[b]Department of Physics, Sinop University, Sinop, Turkey

[c]Department of Physics, Nigde University, Nigde, Turkey

**Abstract**

Nuclear shape transition has been actively studied in the past decade. In particular, the understanding of this phenomenon from a microscopic point of view is of great importance. Because of this reason, many works have been employed to investigate shape phase transition in nuclei within the relativistic and non-relativistic mean field models by examining potential energy curves (PECs). In this paper, by using layered feed-forward neural networks (LFNNs), we have constructed consistent empirical physical formulas (EPFs) for the PECs of $^{38-66}$Ti calculated in Hartree-Fock-Bogoliubov (HFB) method with SLy4 Skyrme forces. It has been seen that the PECs obtained by neural network method are compatible with those of HFB calculations.

**Keywords:** Neural networks, empirical physical formula, potential energy curves, phase transition, Hartree-Fock-Bogoliubov method

E-mails: sakkoyun@cumhuriyet.edu.tr (S.Akkoyun); tbayram@sinop.edu.tr (T. Bayram); sokara@science.ankara.edu.tr (S.O.Kara); nyildiz@cumhuriyet.edu.tr (N.Yildiz)

## 1. Introduction

The study of the structural evolution in atomic nuclei with changing numbers of their neutron and proton constituents dates back to the early days of the nuclear physics. In the last decade, a number of theoretical developments have given insights into, and ways to model, this structural evolution, particularly in transitional regions of rapid change [1, 2]. These breakthroughs involve the concepts of quantum phase transitions (QPTs) and the critical-point symmetries. A new class of symmetries E(5) and X(5) have been suggested to describe shape phase transitions in atomic nuclei by Iachello [3, 4]. The E(5) critical-point symmetry has been found to correspond to the second order transition between U(5) and O(6), while the X(5) critical-point symmetry has been found to correspond to the first order transition between U(5) and SU(3). These symmetries was experimentally identified in the spectrum of $^{134}$Ba [5] and $^{152}$Sm [6].

From theoretical point of view, QPTs have been studied within the Interacting Boson Model (IBM) and the solutions of Bohr-Mottelson differential equations. They are useful representations for describing QPTs in nuclei. Also, phenomenological mean field models (e.g, Hartree-Fock-Bogoliubov (HFB) method [7, 8] and relativistic mean field (RMF) model [9, 10, 11]) have been used to investigate the critical-point nuclei with E(5) or X(5) symmetry in Ref. [12, 13, 14, 15, 16, 17, 18, 19, 20]. In these studies, potential energy curves (PECs) obtained from quadrupole constrained calculations have been used for describing the possible critical-point nuclei. Relatively flat PECs are obtained for critical-point nuclei with E(5) symmetry, while in nuclei with X(5) symmetry, PECs with a bump are obtained. It should be noted, however, that one should go beyond a simple mean field level for a quantitative analysis of QPT in nuclei. For this purpose, some methods have been utilized in Ref. [21, 22, 23, 24]. The application of these methods for a systematic study of QPT in nuclei is at present still very time-consuming. Therefore, the evolution of the PECs along the isotopic or isotonic chains is important and can be used for qualitative understanding of QPTs in nuclei.

In Ref. [19], HFB method with SLy4 Skyrme forces have been employed to investigate ground-state properties of even-even $^{38-66}$Ti isotopes. The calculated binding energies and deformations with the Skyrme force were obtained in good agreement with the available experimental data. In particular, shape evolution of Ti isotopes has been investigated by using calculated PECs to search E(5) symmetry in Ti isotopes, together with the neutron single-particle levels. Particularly $^{46}$Ti has been suggested as to be possible critical-point nuclei with E(5) symmetry.

Recently, neural networks have emerged with successful applications in many fields, obtaining potential energy surfaces [25], studying nuclear mass systematic [26], investigating nucleon separation energies [27], classifying unknown energy levels [28], estimating the density functional theory energy [29], investigating ground state geometries [30], mapping potential energy surfaces [31], determination of beta decay half lives [32] and identifying impact parameter in heavy ion collisions [33]. In this work, borrowing data from our previous work [19], the PECs for $^{38-66}$Ti isotopes as a function of quadrupole deformation parameter ($\beta_2$) were obtained by using layered feed-forward neural networks (LFNNs).

Due to the physical phenomena correlated with potential energy curves (PECs) of the isotopes are characteristically highly non-linear, it may be difficult to construct empirical physical formulas (EPFs) for binding energy functions. By appropriate operations of mathematical analysis, derivation of highly non-linear physical functions for binding energies is of utmost interest. These EPFs would be used for specific purposes in analyzing PECs. We particularly aim to construct explicit mathematical functional form of LFNN-EPFs for PECs. While the PECs were intrinsically non-linear, even so training set LFNN-EPFs successfully fitted these binding energies. Furthermore, testing set LFNN-EPFs consistently predicted the binding energies. That is, the physical laws embedded in the data were extracted by the LFNN-EPFs.

The letter is organized as follows. In Section 2, the HFB method and numerical details are given briefly. In Section 3, details on artificial neural networks are given. The results of this study and discussions are presented in Section 4. Finally, conclusions are given in Section 5.

## 2. HFB Formalism and Numerical Details

In HFB method, many properties of the nuclei can be described in terms of a model of independent particles which move in an average potential. In the HFB formalism, a two-body Hamiltonian of a system of fermions can be interpreted in terms of a set of annihilation and creation operators. The ground state wave function is described as the quasi-particle vacuum and the linear Bogoliubov transformation provides connection between the quasi-particle operators and the original particle operators. The basic building blocks of the HFB method are the density matrix and the pairing tensor, and expectation value of the HFB Hamiltonian could be expressed as energy functional (Details can be found in [8, 34]). In term of Skyrme forces, the HFB energy has the form of local energy density functional contains the sum of the mean field and pairing energy densities. These fields can be calculated in the coordinate space [8, 34].

In this work, input data for construction of empirical formula of the PECs obtained from constrained HFB calculations with SLy4 Skyrme force for Ti isotopes was taken from Ref. [19]. In this reference, HFB equations have been solved by expanding quasi-particle wave functions on a harmonic oscillator basis expressed in coordinate space. For pairing, Lipkin-Nogami method was implemented by performing the HFB calculations with an additional term included in the HF Hamiltonian. Further details on choosing of oscillator bases and parameters can be found in Ref. [19].

## 3. Artificial Neural Networks

Artificial neural networks (ANNs) are known to be very powerful multivariate tools that are used when standard techniques fail to properly take account of the correlation between these variables. The typical goal of the ANN is to get a fast function, which models well the output of complicated and CPU consuming data. Since trained network is very fast and does not use neither much memory nor CPU, ANN is well suited for this task. ANNs offer several advantages, requiring less formal statistical training, ability to detect complex highly non-linear relationships between input and output variables, ability to detect all possible interactions between predictor variables. Another benefit of the ANNs appears in case of existing dataset with a high percentage of missing data.

ANNs are mathematical models that mimic the human brain. They consist of several processing units called neurons which have adaptive synaptic weights [35]. ANNs are also effective tools for pattern recognition. The LFNN which is particular kind of ANN consists of three layers: input, hidden and output (Fig.1). The number of hidden layers can differ, but a single hidden layer is enough for efficient non-linear function approximation [36]. In this study, one input layer with one neuron, one hidden layer with many (*h*) neuron and one output layer with one neuron (*1-h-1*)LFNN topology was used for accurately and reliably prediction of the binding energies for even-even $^{38-66}$Ti isotopes. Analyses were performed for different hidden neuron numbers, $h$= 4, 9 and 14. The total numbers of adjustable weights were 8, 18 and 28.

where $p$ and $r$ are the input and output neuron numbers, respectively.

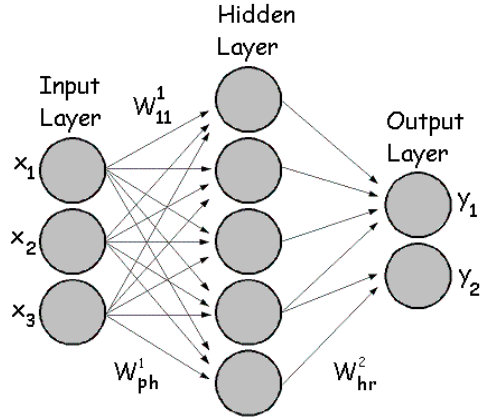

**Fig.1.** Three main layers structure of artificial neural networks

The neuron in the input layer collects the data from environment and transmits via weighted connections to the neurons of hidden layer which is needed to approximate any non-linear function. The hidden neuron activation function can be theoretically any well-behaved non-linear function. In this work, the type of activation function was chosen hyperbolic tangent ($\tanh = (e^x - e^{-x})/(e^x + e^{-x})$) for hidden layer. Finally, the output layer neuron returns the signal after the analysis. Note that, an input layer with single neuron is firmly equivalent to one neuron LFNN with an appropriate activation function. As far as the activation function is analytical, the output is also an analytical function of the input.

Neural network software NeuroSolutions v6.02 was used [37]. The LFNN inputs were quadrupole deformation parameters ($\beta_2$) for $^{38-66}$Ti isotopes and the desired outputs were binding energies. As mentioned before, this data for both training and testing stages were borrowed from our previous work [19]. For all LFNN processing case, whole data were divided into two separate sets, 75% for the training stage and 25% for the testing stage. In the training stage, a back-propagation algorithm with Levenberg-Marquardt [38, 39] for the training of the LFNN was used. By convenient modifications, LFNN modifies its weights until an acceptable error level between predicted and desired outputs is attained. The error function which measures the difference between outputs was *mean square error (MSE)*. Then by using LFNN with final weights, the performance of the network is tested over an unseen data. If the predictions of the testing dataset are good enough, the LFNN is considered to have consistently learned the functional relationship between input and output data [40]. In this study for different *h* numbers, the minimum MSE values were between 0.0005 and 0.0 for training stage and 0.001 and 0.03 for testing stage.

### 3.1 The tangible algorithm for detector response LFNN-EPF construction

In order to construct suitable EPF for highly non-linear PECs, we used one neuron output LFNN vector function $\vec{f}$ as

$$\vec{f}: R^p \to R^r : \vec{f}_k(\vec{x}) = \sum_{j=1}^{h_1} \beta_j G(A_j(\vec{x})), \vec{x} \in R^p, \beta_j \in R, A_j \in A^p, \text{ and } k = 1,...,r \quad (1)$$

where $\vec{w}$ is input to hidden layer weight vector, $\vec{x}$ is the LFNN input vector in Fig. 1, $b$ is the bias weight and $A^p$ is the set of all functions of $R^p \to R$ defined by $A(\vec{x}) = \vec{w}.\vec{x} + b$. However, this equation is not sufficient for the complete construction of the desired non-linear EPF since it gives only the rough structure of the LFNN without generating the final EPF parameters/final LFNN optimal weights. Therefore, in order to obtain the final weight vector $\vec{w}_f$ (consisting of the final components of $w^1$ and $w^2$ of Fig. 1) and the corresponding LFNN output vector function $\vec{f}_{min} = \vec{f}(\vec{w}_f)$ of Eq. (1), we simultaneously used the *MSE values* and Eq. (1). More clearly, given the desired input-output data, $\vec{f}_{min}$ is the network output vector function of Eq. (1) which gives the minimum MSE by a convenient LFNN weight adaptation. Note that, $\vec{f}_{min}$ is the best non-linear estimation vector of the theoretically unknown desired output function $\vec{y}: R^p \to R^r$. In other saying, the unknown vector function $\vec{y}$ is estimated by $\vec{f}_{min}$ which is actually desired non-linear EPF that we aim to eventually obtain. Since $\vec{f}_{min}$ is the highly important quantity, we restate it in Eq. (2).

$\vec{f}_{min}$ of this paper : Given input-output data ($\vec{x}$ *and* $\vec{y}$ samples) and final weight vector $\vec{w}_f$, LFNN $\vec{f}_{min} = \vec{f}(\vec{w}_f)$ is our desired binding energy EPF. In this paper, LFNN input was the quadrupole deformation parameters ($\beta_2$). Desired vector $\vec{y}$ was binding energy. Final details for $\vec{f}_{min}$ of this paper are given Section 3.3. (2)

### 3.3 Final $\vec{f}_{min}$ details

$\vec{f}_{min}$ totally depends on the structure of the network output vector function $\vec{f}$ and the final weight vector $\vec{w}_f$. In Eq. (1), components of the weight are embedded in $A(\vec{x})$ and $\vec{\beta}$. In Eq. (1), $\vec{f}$ depends on the apparent forms of activation and $A$ functions. In this paper, setting $\vec{\beta} = w^2$ of Fig. 1, activation function is non-linear tangent hyperbolic and $A$ is the dot product of $w^1$ and $\vec{x}$ of Fig. 1. So, we can construct explicit form of $\vec{f}$. Afterwards, by minimization of *MSE values*, we finally obtain $\vec{f}_{min} = \vec{f}(\vec{w}_f)$. Now, the concrete LFNN-EPF construction algorithm for non-linear PECs is completed. The actual LFNN-EPFs results are given in Section 4.

### 4. Results and discussions

In figures and text where it suitably applies, the abbreviation *calc* is used for the calculated data obtained by HFB theory. As mentioned in Section 3, the LFNN training and testing set data used in this paper were borrowed from [19]. Note that, the LFNN inputs and outputs used in this paper were explicitly defined in Eq. (1). Inputs were quadrupole deformation parameters ($\beta_2$) and the corresponding outputs were binding energies of the Ti isotopes. The abbreviation *nno* (neural network output) is for both training or testing set results.

## 4.1 Training

For the PECs, the training set *nno* fittings were given in terms of quadrupole deformation parameters ($\beta_2$) versus binding energies of the Ti isotopes obtained from SLy4 Skyrme force (Fig. 2). The LFNN had a single hidden layer with *h=4, 9, 14*. The number of data points belonging to the training stage was about 75% of overall data. It can be clearly seen in Fig. 2, $^{42,50,62}$Ti isotopes which have shell closure with magic neutron numbers (N=20, N=28) and semi-magic number (N=40) were found to be spherical. Also $^{40}$Ti was found to be spherical while $^{38}$Ti has prolate shape. The PECs of $^{44}$Ti seems relatively flat with a small bump which means that it is possible example of β-soft nucleus. The PEC of $^{46}$Ti and $^{48}$Ti is flat from $\beta_2$=-0.2 to $\beta_2$=0.35 and $\beta_2$=-0.2 to $\beta_2$=0.2, respectively. In both of the PECs, the variations of the total binding energies are less than 2 MeV through these $\beta_2$ intervals which implies that the barriers against deformation are so weak. However, the PECs of $^{46}$Ti is much flatter than those of $^{48}$Ti and the PECs of $^{48}$Ti has a small bump in Fig. 2. This means that $^{46}$Ti should be an example of critical-point nuclei with E(5) symmetry, while $^{48}$Ti can be thought as an example of candidate critical-point nuclei with X(5) symmetry. In addition, $^{54-58}$Ti nuclei have flat PECs with a very small bump, while $^{52}$Ti and $^{60}$Ti have flat PECs. It is possible argue that $^{54-58}$Tican be an example of possible critical-point nuclei with X(5) symmetry while $^{52}$Ti and $^{60}$Ti should be candidate for E(5) symmetry.

Moreover in the same figures, one would also concentrate only on comparing specific *nno* fittings with its corresponding *calc* values. In Fig. 2, the *nno* fittings agree exceptionally well with highly non-linear *calc* values. Particularly note that, as principally aimed in this paper, the obtainment of PECs had been successfully made by the LFNN-EPFs.

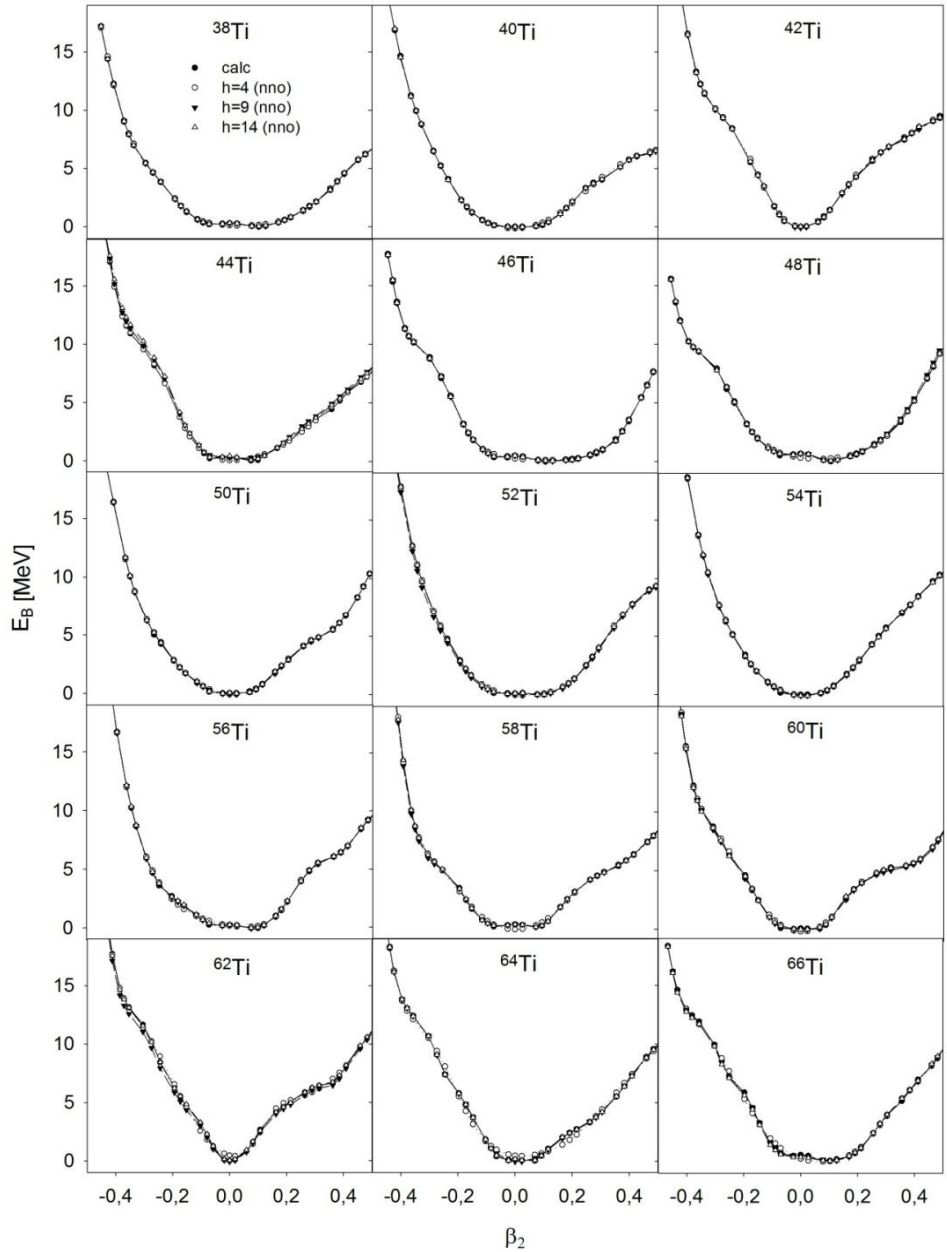

**Fig.2.** HFB calculation (*calc*) with SLy4 Skyrme force and neural network output (*nno*) training set fittings for PECs of different Ti isotopes

*4.2 Testing*

Unless the training set LFNN-EPFs are further tested over the data, these fitted EPFs cannot be used consistently over a desired range of the data. In other words, if the training sets LFNNs well predict previously unseen testing set data, then the LFNN have regarded to have successfully generalized the data, proving consistent estimations. If the estimations are consistent with the testing data values, then the LFNNs can be taken as appropriate LFNN-EPFs. For the PECs, the corresponding testing set *nno* predictions of Fig.2 were given in Fig.3. Naturally, the single hidden layer training set LFNNs with *h=4, 9, 14* which led to Fig. 2 were also used for *nno* testing set predictions. As can be seen in Fig.3, the *nno* predictions agree exceptionally well with highly non-linear *calc* values. This clearly shows that the testing set LFNNs of the quadrupole deformation parameters ($\beta_2$) versus binding energies of the Ti isotopes have consistently generalized the training LFNN fittings. Therefore, LFNNs obtained can be safely used as LFNN-EPFs because the physical law embedded in the quadrupole deformation parameters ($\beta_2$) versus binding energies of the Ti isotopes data has been successfully extracted by the LFNN constructed. Particularly note that, as principally aimed in this paper, the PECs have been successfully made by the LFNN-EPFs.

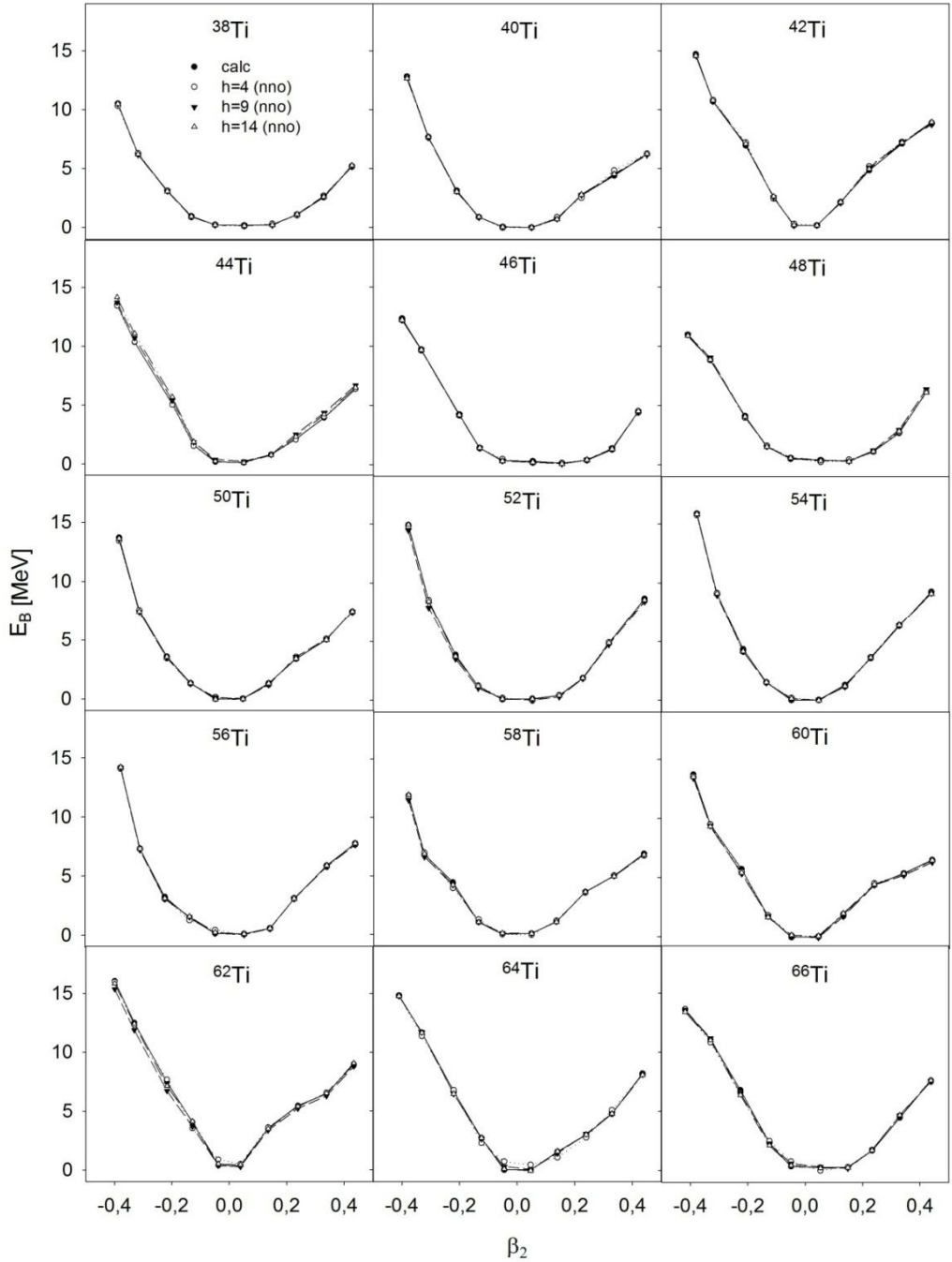

**Fig.3.** HFB calculation (*calc*) with SLy4 Skyrme force and neural network output (*nno*) testing set predictions for PECs of different Ti isotopes

## 5. Conclusions and potential applications

In this paper, based on inputs of the PECs obtained from constrained HFB calculations with SLy4 Skyrme force for $^{38-66}$Ti isotopes, we generated PECs distributions by using artificial neural networks. The PECs of nuclei can provide knowledge for qualitatively understanding of QPTs in nuclei. These distributions can help determination of the nuclear shapes. It is clearly seen that the neural network method, which can be applied very fast, was consistent with the calculated results. The advantage of the ANN method is that it does not need any relationship between input and output data. For highly non-linear binding energies for quadrupole deformation parameters ($\beta_2$), we have novelly constructed consistent empirical physical formula (EPFs) by appropriate LFNNs. The testing set LFNNs of the quadrupole deformation parameters ($\beta_2$) versus binding energies of the Ti isotopes have generalized the training LFNN fittings. Therefore, the testing set LFNNs can be confidently used as LFNN-EPFs.

15[39] D. Marquardt, "An Algorithm for Least-Squares Estimation of Non-linear Parameters," SIAM J. Appl. Math., 1963, Vol. 11, pp. 431–441.

[40] N. Yıldız, "Layered Feedforward Neural Network is Relevant to Empirical Physical Formula Construction: A Theoretical Analysis and Some Simulation Results", Phys. Lett. A 345 (1-3) (2005) 69.